\documentstyle[aps,epsf,multicol]{revtex}
\begin{document}
\draft
\title{The sublattice magnetizations of the spin-$s$ quantum Heisenberg 
antiferromagnets}
\author{Bang-Gui Liu}
\address{Institute of Physics, Chinese Academy of Sciences, 
Beijing 100080, P. R. China}
\date{\today}
\maketitle
\widetext
\begin{abstract}
We study the spin-$s$ quantum Heisenberg antiferromagnetic 
models in a magnon theory free of any unphysical magnon state. 
Because the unphysical magnon states are completely removed
in the magnon Hamiltonians and during the approximation process, we
derive spin-$s$ $(s>\frac 12)$ reliable N\'eel temperature $T_N$ and reasonable
sublattice magnetization unified for $T\leq T_N$.
\end{abstract}


\raggedcolumns
\begin{multicols}{2}
\narrowtext

Quantum Heisenberg antiferromagnetic (QHAFM) models are
well-accepted models for insulating antiferromagnets. 
The insulating cuprates $La_2CuO_4$ and $RBa_2Cu_3O_6$ ($R$ is an 
rare earth element) and their lightdoped samples are typical 
antiferromagnets of spin $1/2$.\cite{cuprate}
Because of their proximity to the high temperature superconductivity
in the doped cuprates, they have absorbed very much attention.
On the other hand, insulating magnetic materials of high spin, such as
the well-known Manganese materials, recently become
important.\cite{hspin} They can be described by the QHAFM models.
But, for the QHAFM models, 
there has not yet existed any reasonable spin-$s$ ($s>\frac 12$)
N\'eel temperature $T_N$ and
sublattice magnetization unified for all temperature $T\leq T_N$.

On the other hand, 
the defining spin operators in the QHAFM models are neither Bose nor
Fermi operators. Many authors tried to transform the spin operators into
the standard Bose (magnon) operators 
since Bloch invented the concept of spin wave.\cite{bloch}
On the experimental side authors tend to make use of spin wave (magnon)
theory to explain their experimental results.\cite{exp}
Therefore, magnon (or spin wave) theory is a very powerful
approach to the QHAFM models.
In this approach one makes use of Holstein-Primakoff\cite{hp} or 
Dyson\cite{dyson} transformation to transform the original spin 
Hamiltonians into magnon Hamiltonians. On a single site, 
the magnon state space is infinitely dimensional, but
there are only $2s+1$ physical spin states. There exist infinite extra 
unphysical magnon states in any magnon Hamiltonian.
In removing effect of the unphysical states, many methods were 
proposed,\cite{mattis,nlsw,kubo,oguchi,takahashi,other} 
but the problem of the unphysical magnon states remained unsolved.
In a magnon theory free of the unphysical states,\cite{liuczyc}
all of the unphysical states were put infinitely high in energy
and their effect was removed in the magnon Hamiltonians and during
the approximation process. In the case of spin $1/2$, reliable 
ferromagnetic magnetization and antiferromagnetic sublattice 
magnetization were obtained under a simple but sensible approximation.

In this Letter we study the ($s\geq 1$) quantum 
Heisenberg antiferromagnetic models 
in the magnon theory free of the unphysical states. 
In the way similar to those in the previous works,\cite{liuczyc}
we remove effect of the unphysical magnon states in the magnon
Hamiltonians and during the approximation process, and thereby 
derive for the first time reliable analytical
N\'eel temperature $T_N$ and sublattice magnetization,
being reasonable from zero to $T_N$, 
of the quantum Heisenberg antiferromagnetic
model of any spin ($s\geq 1$) in three dimensions. 
The zero-temperature normalized sublattice magnetization $M_0/s$
increases monotonously with spin $s$. The reduced sublattice 
magnetization $M/M_0$ decreases monotonously with spin $s$ 
increasing at any given reduced temperature $T/T_N$ 
between $0$ and $1$. Our sublattice
magnetizations are substantial improvements on those of the existing
nonlinear spin wave (magnon) theories, which became unreasonable at high 
temperature. Our N\'eel temperature in the spin-$\frac 12$ case 
is only 4\% larger than the series expansion result.\cite{d}

The standard spin-$s$ QHAFM Hamiltonian is defined by
\begin{equation}
H=\sum_{\langle ij\rangle }J_{ij}\left[\frac 12
(S_i^{+}S_j^{-}+S_i^{-}S_j^{+})+S_i^zS_j^z\right]. 
\end{equation}
The summation is over the nearest neighbor sites of the lattice. 
The exchange constant $J_{ij}$ is positive and, in general, is 
direction dependent to describe real space anisotropy.
We suppose our lattice can be divided into two sublattices so
that the nearest neighbors of any site of one sublattice belong
to the other sublattice. Making a $\pi$ rotation in the spin 
space of one of the two sublattices, the Hamiltonian (1) becomes
\begin{equation}
H=\sum_{\langle ij\rangle }J_{ij}\left[\frac 12
(S_i^{+}S_j^{+}+S_i^{-}S_j^{-})-S_i^zS_j^z\right].  \label{Heis}
\end{equation}
To transform the spin operators into the magnon operators,
We choose Holstein-Primakoff (HP) transformation: 
\begin{equation}\label{mt}
S_i^{-}=a_i^{\dagger }\sqrt{2s-n_i},~~S_i^{+}=\sqrt{2s-n_i}a_i,~~
S_i^z=s-n_i\label{HP},
\end{equation}
where the magnon number operator $n_i$ is defined by $n_i=a_i^{\dagger }a_i$. 
The magnon operators $a$ are standard Bose operators.
After substituting the magnon transformation (\ref{mt}) into the Hamiltonian
(\ref{Heis}), we obtain the HP magnon Hamiltonian: 
\begin{equation}\label{magnon}
H_m=\sum_{\langle ij\rangle }J_{ij}\left[\frac 12
(a_i^{\dagger }a_j^{\dagger }A_iA_j+
{\rm h.c.}) -n_in_j\right]+\sum_i\epsilon a_i^{\dagger }a_i+H_0
\end{equation}
where $H_0=-\frac 12JZs^2N$, $\epsilon =JZs$, and $A_i=\sqrt{2s-n_i}$. 

The magnon Hamiltonian (\ref{magnon}) is different from the original
spin Hamiltonian (\ref{Heis}) because the magnon operator on a single
site has infinite states. For spin $s$ we have only
$2s+1$ physical spin states on a single site. To remove effect of the
infinite unphysical magnon states, we introduce an infinite-$U$ 
term:\cite{liuczyc}
\begin{equation}
H_U=\sum_i\frac U{(2s+1)!}a_i^{\dagger (2s+1)}a_i^{(2s+1)},~~~U\rightarrow
\infty  \label{HU}
\end{equation}
into the HP magnon Hamiltonian.
As a result, our total Hamiltonian is defined by $H^{\prime }=H_m+H_U$. 
The extra unphysical states are raised infinitely high in energy by
the $U$ term and 
actually decouple from the $2s+1$ physical states. 
The magnon square root $A_i=\sqrt{2s-n_i}$ in the Hamiltonian
can be expanded in terms of $a_i^{\dagger m}a_i^m$. 
The $U$ term makes the unphysical onsite
correlation $\left\langle a_i^{\dagger m}a_i^m\right\rangle $ 
($m\geq 2s+1$) exactly equivalent to zero. 
So that the expansion of $\sqrt{2s-n_i}$ is accurately
truncated into a sum of $2s+1$ magnon operator product terms: 
\begin{equation}
 A_i=\sum_{m=0}^{2s}(-1)^mC_m\frac 1{m!}
a_i^{\dagger m}a_i^m 
\label{sqrt}
\end{equation}
\begin{equation}\label{cm}
\displaystyle C_m=\sum_i^m(-1)^i\frac{m!}{i!(m-i)!}\sqrt{2s-i}.
\end{equation}
Substituting the $A_i$ expression (\ref{sqrt}) into the magnon 
Hamiltonian (\ref{magnon}), our total Hamiltonian becomes a
finite sum of rational magnon operator products.
Therefore, the infinite-$U$ term $H_U$ removes effect of the 
unphysical magnon states and at the same time simplify the
magnon Hamiltonian.\cite{liuczyc}

Being similar to the previous work,\cite{liuczyc} we work
in the framework of Green function method and the approach of
equation of motion. We define $\langle\langle X|Y\rangle\rangle$ to be
the Green function of the operators $X$ and $Y$ in Zubarev notation.
The equation of motion of operator $X$ is defined to be 
$i\frac d{dt}X=[X,H']$, or $zX=[X,H']$ in the frequency ($z$) space.
Since the magnon number is not conserved for the AFM models, we need
two kinds of Green functions.
Intuitively, we have $\langle\langle a_i|a^{\dagger}_j\rangle\rangle$
and $\langle\langle a^{\dagger}_i|a^{\dagger}_j\rangle\rangle$.
But they cannot compose a closed set of equations of motion because
of the infinite $U$ term in the magnon Hamiltonian.
On-site multimagnon Green functions inevitably appear due to
the $U$ term. We need a series of the on-site multimagnon 
Green functions to construct some composite Green functions
so that their equations of motion are closed and contain terms of
at most $1/U$ order.\cite{liuczyc} For this reason, 
we construct the two series of Green functions:
$\langle\langle B_i|D_j^m\rangle\rangle$ and
$\langle\langle B_i^{\dagger}|D_j^m\rangle\rangle$
($m=1,2,3,\cdot\cdot\cdot,2s$),
where $D_j^m=a_j^{\dagger m}a_j^{\dagger m-1}$ and $B_i=\bar{A}_ia_i$. 
Here the operator $\bar{A}_i$ is equivalent to $A_i$ with
the $a_i^{\dagger 2s}a_i^{2s}$ term removed.
The Green functions satisfy the equations of motion
\begin{equation}\label{eom1}
(z-\epsilon)\langle\langle B_i|D_j^m\rangle\rangle
=W^m_i\delta_{ij}
+\sum_lJ_{il}\langle\langle T_{il}|D_j^m\rangle\rangle
\end{equation}
and
\begin{equation}\label{eom2}
(z+\epsilon)\langle\langle B_i^{\dagger}|D_j^m\rangle\rangle
=\bar{W}^m_i\delta_{ij}
-\sum_lJ_{il}\langle\langle T^{\dagger}_{il}|D_j^m\rangle\rangle,
\end{equation}
where $W^m_i=\langle[A_ia_i,D_i^m]\rangle$, 
$\bar{W}^m_i=\langle[a^{\dagger}_iA^{\dagger}_i,D_i^m]\rangle$, and
$T_{il}=(s-n_i)a^{\dagger}_lA_l-n_lA_ia_i$.
In Eqs. (\ref{eom1}) and (\ref{eom2}), other terms of $1/U$ order 
or smaller terms are neglected since they are zero when $U$ tends to
infinity.
We make the on-site approximation 
$n_iX_l=\langle n_i\rangle X_l=\langle n\rangle X_l$ 
for the operator $n_iX_l$ only if $i\not= l$.
Under this approximation, the equations of motion reduce to
\begin{equation}\label{em1}
(z-\gamma)\langle\langle B_i|D_j^m\rangle\rangle=
W^m_i\delta_{ij}+\sum_lJ_{il}(s-n)
\langle\langle B_l^{\dagger}|D_j^m\rangle\rangle
\end{equation}
and
\begin{equation}\label{em2}
(z+\gamma)\langle\langle B_i^{\dagger}|D_j^m\rangle\rangle=
-\sum_lJ_{il}(s-n)\langle\langle B_l|D_j^m\rangle\rangle,
\end{equation}
where $\gamma=JZ(s-n)$. The parameter $J$ is the largest one of
the exchange $J_{ij}$ and the effective coordination number 
$Z$ is defined by $Z=\sum_{j(i)}J_{ij}/J$, where the $j$ 
summation is made over all nearest neighbor sites of site $i$.
It is not difficult to solve the above equations by transforming the
Green functions into the $k$-space. The resultant Green functions
in the $k$-space read:
\begin{equation}\label{em11}
\langle\langle B_k|D_k^m\rangle\rangle=\frac
{W^m[z+JZ(s-n)]}{z^2-J^2Z^2(s-n)^2(1-r_k^2)}
\end{equation}
and
\begin{equation}\label{em22}
\langle\langle B_{-k}^{\dagger}|D_k^m\rangle\rangle=\frac
{-W^mJZ(s-n)r_k}{z^2-J^2Z^2(s-n)^2(1-r_k^2)}.
\end{equation}
The spectrum function $r_k$ is defined by 
$r_k=(2/JZ)\sum_{l(j)}J_{jl}\exp (iR_l-iR_j)$.
It is obvious that the magnon correlation functions 
$\langle D^m_iB_i\rangle=V^m_i$ and $W^m_i$
both are site independent. Therefore we obtain the self-consistent 
equations
\begin{equation}\label{V}
V^m=W^m\psi
\end{equation}
and
\begin{equation}\label{psi}
\psi =\frac 1N\sum_k\frac 12\left[\frac{1}{\sqrt{1-r^2_k}}
\coth \left(\frac{\beta\gamma}{2} \sqrt{1-r^2_k}\right)-1\right].
\end{equation}
For spin $s$, we have only $2s$ non-zero on-site correlation functions:
$p_n=\langle P_{ni}\rangle$ $(1\leq n\leq 2s)$, 
where the on-site multimagnon operator 
$P_{ni}$ is defined by $P_{ni}=\frac{1}{n!}a^{\dagger n}_ia^n_i$.
The unphysical correlation functions $P_n$ ($n\geq 2s+1$) are exactly 
equivalent to zero. 
The self-consistent equation (\ref{V}) can be solved by
\begin{equation} \label{pn}
p_n=\frac{\psi^n}{(1+\psi)^{2s+1}-\psi^{2s+1}}
\displaystyle\sum^{2s-n}_{k=0}C^{2s+1}_{k}\psi^k,
\end{equation}
where $n\le 2s$ and the coefficient $C^n_m=n!/[m!(n-m)!]$ is the binomial 
coefficient. 
Since $S^z_i=s-n_i$ and $n=p_1$, we derive the following spin magnetization:
\begin{equation}\label{s3}
\left\langle S^z\right\rangle 
=s-\psi+\frac{(2s+1)\psi^{2s+1}}{(1+\psi)^{2s+1}-\psi^{2s+1}}.
\end{equation}
At very low temperature, $\psi$ can be expanded as $\psi=\psi_0+cT^2$
($c$ is a positive constant.),
where $\psi_0=(\xi -1)/2$ and $\xi=\frac{1}{N}\sum_k1/\sqrt{1-r^2_k}$.
As a result, we derive $\left\langle S^z\right\rangle=S^z_0-c_1T^2$
($c_1$ is also a positive constant.).
$S^z_0$ is given by the right-hand side of formula (\ref{s3}) with
$\psi$ replaced by $\psi_0$.
The N\'eel temperature $T_N$ is given by 
\begin{equation}
T_N=\frac{JZ}{3\eta} s(s+1)  \label{tcs},
\end{equation}
where the constant $\eta$ is given by $\eta=\frac 1N\sum_k 1/(1-r^2_k)$.
When temperature approaches to the N\'eel temperature $T_N$, the 
sublattice magnetization $\left\langle S^z\right\rangle$ is given by
the asymptotic expression:
\begin{equation}
\left\langle S^z\right\rangle=\alpha\sqrt{1-\frac T {T_N}}.
\end{equation}
The positive constant $\alpha$ is defined by 
$\alpha=\frac 23 s(s+1)[1+\frac{\eta}3+\frac 2{15}(2s^2+2s-9)]^{-1/2}$.
In one dimension, the integrations $\eta$ and 
$\xi$ both diverge. There is no
antiferromagnetic order at any temperature. In two dimensions
the integration $\xi$ converges, but the integration $\eta$ diverges.
There is antiferromagnetic order only at zero temperature.
These are consistent with the Mermin-Wagner theorem.\cite{mw}

At general temperature, one cannot derive any analytical expression.
We present some digital results in three-dimensional (3D) simple-cubic
lattice.
In Fig. 1 we show the sublattice magnetizations of spins $1/2$, $1$, 
$3/2$, and $2$ as functions of temperature. For convenience, we
normalize the saturation magnetizations into one, that is, we divide the 
magnetizations by the corresponding spin $s$.
The normalized sublattice magnetizations increase with spin $s$ 
increasing at zero temperature. 
For comparison with the results of the existing spin wave (magnon) theories,
we also present the sublattice magnetizations of a typical version of
the existing nonlinear spin wave (magnon) theories (NLSW). 
It is obvious 
that the NLSW magnetizations are unreasonable at high temperature,
especially near the N\'eel temperatures. In fact, all of the existing
NLSW theories yielded double-valued magnetizations at high 
temperature.\cite{matt,mmm}
Therefore the existing spin wave (magnon) theories work at most at low 
temperature, and our magnon theory free of unphysical states
works well in whole temperature region.
%%%%%%%%%%%%%%%%%%%%%%%%%%%%%%%
\vspace{-1.8cm}
\begin{figure}[tbp]
\epsfxsize=0.45\textwidth
\epsfbox{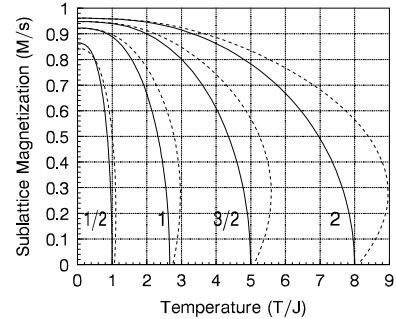}
\caption{Sublattice magnetizations (divided by $s$) 
as functions of temperature (in unit of $J$) for spin
$1/2$ (see Ref 13), $1$, $3/2$, and $2$. The solid lines are the results of 
the magnon theory free of any unphysical states. The dashed lines 
are results from other magnon (spin wave) theories (Ref 16). The latter 
are not reasonable at high temperature (Refs 16, 17). }
\label{fig1}
\end{figure}
\vspace{-1.8cm}
%%%%%%%%%%%%%%%%%%%%%%%%%%%%%%%
%%%%%%%%%%%%%%%%%%%%%%%%%%%%%%%%%
\begin{figure}[tbp]
\epsfxsize=0.45\textwidth
\epsfbox{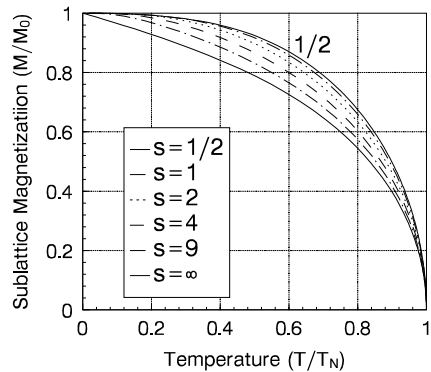}
\caption{The reduced sublattice magnetizations ($M/M_0$) as functions
of reduced temperature ($T/T_N$) for spin $1/2$ (see Ref 13), 
$1$, $2$, $4$, $9$, 
and $\infty$. The reduced sublattice magnetization decreases with 
reduced temperature increasing for a given spin $s$. It also decreases 
with spin $s$ increasing at a given reduced temperature.}
\label{fig2}
\end{figure}
%%%%%%%%%%%%%%%%%%%%%%%%%%%%%%%%%%

To compare the magnetizations of different spins, we show the reduced
sublattice magnetizations $M/M_0$ of spins $1/2$, $1$, $2$, $4$, $9$, 
and $\infty$ as functions of the reduced temperature $T/T_N$.
Near the N\'eel temperature, the reduced magnetization behaves as
$\sqrt{1-T/T_N}$ for any spin $s$. At low temperature, the reduced
sublattice magnetization can be expanded as $1-c_2 T^2$ ($c_2$ is a
positive constant.). For very
large spin $s$, the $T^2$ behavior holds within a temperature 
region of the width $s$. But the N\'eel temperature is proportional
to $s^2$. As a function of the reduced temperature $t=T/T_N$, the 
the $t$ region of square behavior shrinks to zero when spin $s$
tends to infinity. Therefore, for spin $\infty$, the reduced 
magnetization is linear in the reduced temperature at $T/T_N=t\sim 0$.
It is very interesting that the reduced 
magnetization monotonously decreases with spin $s$ 
at any given reduced temperature $t$ between $0$ and $1$.

The above results are obtained under the postulation of isotropy
($J_{ij}=J$) in the 3D simple cubic lattice.
In fact, the approach and the on-site approximation can
be successfully applied to the cases of anisotropic exchange
constants where $J_{ij}$ is direction dependent. 
In the cases of the cuprates, one has an in-plane exchange
constant $J$ and an interplane exchange constant $\delta J$ 
($\delta$ is small.).\cite{cuprate,liu} For line-like
materials, one has an on-line exchange constant $J$ and an interline
constant $\delta J$ with small $\delta$.

The method and approximation process we make use of in this paper are
similar to the previous work on the quantum Heisenberg spin-$\frac 12$ 
ferromagnetic models and spin-$\frac 12$ antiferromagnetic 
model.\cite{liuczyc}
In those cases, our magnetizations and sublattice magnetization were
equivalent to the reasonable results which had been
obtained by means of the direct spin-operator methods.\cite{bogo,tahir,pu}
In the case of spin 1/2, we obtain $T_N=0.989J$ in the simple cubic 
lattice, being only 4\% larger than the series expansion result
$T_N=0.951J$.\cite{d}
Therefore, the reliability of our spin-$s$ sublattice magnetizations
and N\'eel temperatures
in this paper can be justified by the equivalence in the ferromagnetic 
cases and the comparison with the accurate series expansion 
result.\cite{d}

In summary, we have studied the quantum Heisenberg antiferromagnetic 
models of spin $s$ in a magnon theory free of the unphysical states. 
Effect of the unphysical states is completely removed in the magnon
Hamiltonians and during the approximation process. We thereby
derive reliable analytical the N\'eel temperatures and sublattice magnetizations 
unified for all temperatures $T\leq T_N$
and  obtain the multimagnon correlation
functions for any spin.
At any given reduced temperature, the reduced sublattice 
magnetization monotonously decreases with spin $s$ increasing.
Our sublattice magnetizations are substantial improvements on
those of the existing spin wave (magnon) theories. 
Our N\'eel temperature in the case of spin $1/2$ is only 4\% larger
than the series expansion result.

\section*{Acknowledgment}
The author appreciatively acknowledges that this research is partially supported 
by Chinese Nature Science Foundation, by Chinese Ministry of Science \& Technology
under the State Key Project of Basic Research on Rare Earth,
and by Chinese Ministry of  Education.

\end{multicols}


\begin{references} 
\bibitem{cuprate}  A. P. Kampf, Physics Reports {\bf 249}, 219 (1994);
M. A. Kastner, R. J. Birgeneau, G, Shirane, and Y. Endoh,
Rev. Mod. Phys. {\bf 70}, 897 (1998);
T. E. Mason, in {\it Handbook on the Physics and Chemistry of Rare Earths},
Special Volumes on High Temperature Rare Earth Superconductors, 1998,
Editors: K. A. Gschneidner Jr., L. Eyring, and M. B. Maple.

\bibitem{hspin} {\it e. g.,} 
R. J. Birgeneau, H. J. Guggenheim, and G. Shirane,
   Phys. Rev. B {\bf 1}, 2211 (1970);
   % Neutron scattering of high-spin materials: K2NiF4 Rb2MnF4 Rb2FeF4
M. Hase, I. Terasaki, and K. Uchinokura, Phys. Rev. Letts.
   {\bf 70}, 3651 (1993);  
   % CuGeO3 series
S. Jin, T. H. Tiefel, M. McCormack, R. A. Fastnacht, R. Ramesh, 
   and L. H. Chen, Science {\bf 264}, 413 (1994);  
   % Mn CMR
M. Greven, R. J. Birgeneau, Y. Endoh, M. Kastner, B. Keimer,
   M. Matsuda, G. Shirane, and T. R. Thurston, 
   Phys. Rev. Letts. {\bf 72}, 1096 (1994);
   % Neutron scatterring: K2NiF4
K. Ueda, H. Tabata, and T. Kawai, Science {\bf 280}, 1064 (1998).
   % LaFeO3-LaCrO3 superlattice

\bibitem{bloch}  F. Bloch, Z. Phys. {\bf 61}, 206 (1930); 74, 295 (1932).

\bibitem{exp}  {\it e.g.} 
B. Keimer, N. Belk, R. J. Birgeneau, A. Cassanho, C. Y. Chen, 
   M. Greven, M. Kastner, M. Aharony, Y. Endoh, R. Erwin, and 
   G. Shirane, Phys. Rev. B {\bf 46}, 14034 (1992);
S. M. Hayden, G. Aeppli, H. A. Mook, T. G. Perring, T. E. Mason, 
   S-W. Cheong, and Z. Fisk, Phys. Rev. Letts. {\bf 76}, 1344 (1996).
% experimental papers that emphasized spin wave theory

\bibitem{hp}  T. Holstein and H. Primakoff, Phys. Rev. {\bf 58}, 1098 (1940).

\bibitem{dyson}  F. J. Dyson, Phys. Rev. {\bf 102}, 1217 (1956), 1230 (1956).

\bibitem{mattis}  D. C. Mattis, {\it The theory of magnetism 
$I$ (Statics and
dynamics) (2nd corrected printing)}, Springer-Verlag 1988; 
{\it The theory of
magnetism $II$ (Thermodynamics and statistical mechanics)}, 
Springer-Verlag 1985.

\bibitem{nlsw}  M. Marshall and S. W. Lovesey, 
{\it Theory of Thermal Neutron
Scattering}, Oxford Clarendon Press, 1971.

\bibitem{kubo}  R. Kubo, Phys. Rev. {\bf 87}, 568 (1952).

\bibitem{oguchi} T. Oguchi, Phys. Rev. {\bf 117}, 117 (1960).

%\bibitem{ld} P. Lindg\aa{}rd and O. Danielsen, J. Phys.
%C {\bf 7}, 1523 (1974).

\bibitem{takahashi}  M. Takahashi, Phys. Rev. B {\bf 42}, 766  (1990); 
{\bf 40}, 1494 (1989).
% modified spin wave theory, FM, 2D

\bibitem{other} For other nonlinear spin wave theories see, {\it e. g.} 
 J. Igarashi, Phys. Rev. B {\bf 46}, 10763 (1992);
% NL spin wave theory by HP transformation, 2D
% Zc=1.1794; S0=0.3069
C. M. Canali, S. M. Girvin, and M. Wallin, Phys. Rev. B {\bf 45},
 10131 (1992).
% NL spin wave theory by Dyson transformation, 2D

\bibitem{liuczyc}  B.-G. Liu and G. Czycholl, 
J. Phys. Condensed Matter {\bf 9}, 5449 (1997); 
{\it The 7th Asia Pacific Physical Conference (7APPC)},
Beijing, P. R. China, 19-23 August 1997 (page 119 of The proceedings)

\bibitem{d} D. Jou and H Chen, Phys. Lett. {\bf 45A}, 239 (1973).
% series expansion result for spin 1/2 AFM

\bibitem{mw}  N. D. Mermin and H. Wagner, Phys. Rev. Lett. 
{\bf 17}, 1133  (1966).
% no phase transition at finite T in 2D

\bibitem{matt} See Mattis's books \cite{mattis} for detailed analysis.

\bibitem{mmm} The unreasonable double-valued spin magnetizations 
at high temperature are shared, to our knowledge, by all existing 
nonlinear spin wave (magnon) theories, such as those in 
references\cite{nlsw,kubo,oguchi,takahashi,other}.

\bibitem{liu} B.-G. Liu, Phys. Rev. B {\bf 41}, 9563 (1990).

\bibitem{bogo}  N. N. Bogoliubov and S. V. Tyablikov, 
Dokl. Akad. Nauk. {\bf 126}, 53 (1959).

\bibitem{tahir}  R. A. Tahir-Kheli and D. ter Haar, 
Phys. Rev. {\bf 127}, 88 (1962).

\bibitem{pu} F.-C. Pu, Sov. Phys.-Dokl. {\bf 5}, 128 (1960).

\end{references}
\end{document}